\documentclass[pre,twocolumn]{revtex4}
\usepackage[english]{babel}
\usepackage{amsmath}
\usepackage{amssymb}
\usepackage{epsfig}

\begin{document}
\title{Instabilities of a Filled Vortex in a Two-Component Bose-Einstein Condensate}
\author{Victor\,P.\,Ruban}
\email{ruban@itp.ac.ru}
\affiliation{Landau Institute for Theoretical Physics RAS,
Chernogolovka, Moscow region, 142432 Russia}

\date{\today}

\begin{abstract}
A two-component Bose-Einstein condensate of cold atoms 
with a strong intercomponent repulsion leading
to the spatial separation of the components has been 
numerically studied. Configurations with a multiple
quantized vortex in one component, where the vortex 
core is filled with the other component, are considered.
The effective radius of the core can exceed the width 
of the transition layer between components, and then an
analogy with a filled cylindrical vortex in the classical
hydrodynamics of two immiscible ideal fluids appears.
This analogy allows one to analyze the longitudinal 
``sausage'' instability and the transverse instability of the
filled vortex in the condensate caused by the ``tangential 
discontinuity,'' as well as the stable regime in the
parametric gap between them. The presence of long-lived 
coherent structures formed in some cases at the nonlinear 
stages of both instabilities is numerically discovered.
\end{abstract}

\maketitle

In the theory of Bose-Einstein condensates of cold
atoms, multicomponent models attract considerable
attention. In particular, two gases of different chemical
elements or of atoms of a single element but in two different 
internal quantum states can exist simultaneously [1-5]. 
Stationary configurations, instabilities,
and nonlinear dynamics of such systems are very rich
[6-17], even in comparison with all the variety of
properties and modes that were found in one-component 
condensates [18, 19] (where the quantized vortices 
alone constituted a separate line of research [20-28]).

In particular, very interesting coherent structures
are filled vortices, when a quantized vortex of multiplicity 
$Q$ is present in one of the condensate components, 
and the core of this vortex is a potential well for another 
(``bright'') component [29-31]. The equilibrium 
profile of the well is determined self-consistently
and may differ significantly from the ``empty'' vortex
profile in a one-component condensate. The main
difference is a much larger core width. The linear stability 
of such three-dimensional configurations was
studied in Ref.[31], where a number of unstable modes
were discovered and numerical examples of transitions
to nonlinear regimes were given. In addition, filled
vortices with sufficiently large values $Q\sim 10$--$30$ 
in trapped condensates were simulated numerically to
demonstrate the quantum Kelvin-Helmholtz instability 
in finite systems [12,13]. However, the problem
has not yet been studied comprehensively. The aim of
this work is a simplified, in comparison to Ref.[31], consideration 
of the two main instabilities: the longitudinal ``sausage'' 
instability [32] and the transverse instability 
(of the Kelvin-Helmholtz type). It is demonstrated 
that both instabilities are controlled by one parameter, 
and that a stable region between them exists at moderate $Q$ values.

We start with general comments. Dimensionless equations
of motion for the wavefunctions $A({\bf r},t)$ and  $B({\bf r},t)$
have the form of coupled Gross-Pitaevskii equations [1,3]
\begin{eqnarray}
&&i\dot A=-\frac{1}{2   }\nabla^2 A+\left(V_1+g_{11}|A|^2+g_{12}|B|^2\right)A,
\label{GP1}\\
&&i\dot B=-\frac{1}{2\nu}\nabla^2 B+\left(V_2+g_{12}|A|^2+g_{22}|B|^2\right)B,
\label{GP2}
\end{eqnarray}
where $\nu=M_2/M_1$ is the ratio of atomic masses, $V_\alpha({\bf r},t)$
are external potentials, and $g_{\alpha\beta}$ is a symmetric matrix
of nonlinear interactions. The case of positive $g_{\alpha\beta}$ will
be of interest. Without loss of generality, it can be
assumed that $g_{11}=\kappa$ and $g_{22}=1/\kappa$, so $g_{11}g_{22}=1$ and
the dynamical system given by Eqs.(1) and (2) is
characterized by only three essential dimensionless
parameters (excluding external potentials) $\nu$, $\kappa$,  and
$g=g_{12}-1$. However, this relatively simple model is
applicable only in the limit of zero temperature and
cannot describe any finite-temperature effects. For
comparison, the equations of motion for, e.g., $^3$He
[33], where filled vortices (with a more complex structure 
than those considered here; see [34]) are also possible, 
are more complex (and thermodynamics is of great importance).

When repulsion $g>0$ between components prevails, 
the condensate components tend to separate spatially [6, 7] 
(therefore, $g$ can be called the segregation 
parameter). In particular, in the absence of external 
forces, a transition layer is formed in equilibrium,
which is a stationary one-dimensional solution of the
system of Eqs.(1) and (2)
$$A=a(x)\exp(-i\mu\sqrt{g_{11}}t),\quad B=b(x)\exp(-i\mu\sqrt{g_{22}}t),$$
where $a(x)$ and $b(x)$ are real-valued functions. Certain
additional energy associated with this layer is the
effective surface tension [7, 11]
\begin{eqnarray}
\sigma&=&\mbox{min}\int_{-\infty}^{+\infty}
\Big[ a'^2/2 +b'^2/(2\nu)+g a^2 b^2
\nonumber\\
&&\qquad +(\sqrt{g_{11}}a^2+\sqrt{g_{22}}b^2-\mu)^2/2\Big]dx.
\end{eqnarray}
According to the Maupertuis principle of classical
mechanics, the same quantity can be represented as
\begin{eqnarray}
\sigma&=&\mbox{min}\int_{(a_1,0)}^{(0,b_2)}\sqrt{[(da)^2+(db)^2/\nu]}
\nonumber\\
&&\qquad \times\sqrt{2g a^2 b^2+(\sqrt{g_{11}}a^2+\sqrt{g_{22}}b^2-\mu)^2},
\end{eqnarray}
where the integral is taken along an arbitrary curve [in the $(a,b)$] plane], 
starting at the point ($a_1=\mu^{1/2}g_{11}^{-1/4}$, $b_1=0$)
and ending at the point ($a_2=0$, $b_2=\mu^{1/2}g_{22}^{-1/4}$).
The parameter $\mu$ can be completely taken out as a multiplier  $\mu^{3/2}$
by replacing $(a,b)=\sqrt{\mu}(\alpha,\beta)$. In the case of
small values of $g\ll 1$ , the optimal trajectory passes
near the ellipse $\sqrt{g_{11}}\alpha^2+\sqrt{g_{22}}\beta^2=1$, 
and $\sigma$ is about $\mu^{3/2}\sqrt{g}$ (a more precise expression 
can be found in [11]). The width of the transition layer between the
two components can be estimated as $w\sim 1/\sqrt{g\mu}$.

The presence of surface tension makes the large-scale 
dynamics of the interface in a segregated binary
condensate similar to the dynamics of bubbles in the
classical mechanics of immiscible ideal fluids [15-17].
The flow is potential inside each of the components,
and the entire vorticity of the velocity field is concentrated 
at the interface. In this sense, the bubble boundary 
is a vortex sheet resembling vortex sheets in $^3$He-$A$
[35] in some cases. However, the analogy with classical 
hydrodynamics does not always work; for example,
the equilibrium states of rotating binary condensates
have a complex ``fine'' structure [9, 10, 14].

Consider filled vortices. Let $A$ and $B$ be the vortex
and bright components, respectively. Depending on
the amount of the bright component, the vortex can be
in different modes. If the amount of the bright component 
is so small that the vortex core radius $R$ differs
slightly from the transition layer thickness $w$, the vortex 
continues to be effectively a one-dimensional
object in the dynamic sense. The inertia of a filled vortex, 
as well as its increased thickness, noticeably modifies 
the dynamics in comparison with an empty vortex, 
leading to the appearance of instabilities in a number 
of cases (this mode will be discussed elsewhere).

We consider the effects that occur at $R\gg w$, when
the degrees of freedom associated with the difference
in the cross-sectional shape from the equilibrium circular 
one are excited in the system. In this case, a two-dimensional 
surface of a distorted cylinder separating
two condensates appears instead of a one-dimensional
vortex filament. The typical velocity of motion is significantly 
less than the speed of sound. Therefore,
there is an analogy with a filled cylindrical vortex in
the classical hydrodynamics of two immiscible ideal
fluids (with constant densities $\rho_{\rm in}=\nu\mu/\sqrt{g_{22}}$ and
$\rho_{\rm out}=\mu/\sqrt{g_{11}}$, surface tension $\sigma$, and circulation
$\Gamma=2\pi\tilde\Gamma$). In a stationary state, the inner fluid is at
rest, and the outer one has an azimuthal velocity
$v_\phi=\tilde\Gamma/\sqrt{x^2+y^2}$.

For such a classical filled vortex, two types of instabilities are known, 
depending on the parameter $\Xi=\sigma R/(\rho_{\rm out}\tilde\Gamma^2)$. 
First, at $\Xi>1$, a three-dimensional longitudinal sausage instability develops [32]. 
Its origin is easily understood by writing the effective potential 
energy ${\cal U}$ of axisymmetric motions (the sum of the
surface energy and the kinetic energy of the azimuthal
flow in the outer region) in terms of the canonical variable 
$S(z)=r^2(z)/2$, proportional to the cross-sectional area:
\begin{equation}
{\cal U}\{S(z)\}=\pi\int\Big[2\sigma\sqrt{2S+S'^2} 
-\rho_{\rm out}\tilde\Gamma^2\ln(\sqrt{2S}/R)\Big]dz.
\label{U}
\end{equation}
This functional in the quadratic approximation in the
deviation $s=(S-R^2/2)$ has a negative coefficient at
sufficiently low wavenumbers $k_z$, exactly for $\Xi>1$.

Second, since there is a tangential discontinuity in
the flow velocity at the vortex boundary, the Kelvin-Helmholtz 
instability is possible. Consideration of linearized 
conservative equations describing only two-dimensional 
small perturbations of the cross section of a classical 
filled vortex leads to the dispersion law for azimuthal modes in the form
\begin{eqnarray}
\omega_m=\frac{\tilde\Gamma}{R^2(1\!+\!\rho)}\qquad\qquad\qquad\qquad&&\nonumber\\
\times\Big\{m+\!\!\sqrt{(1\!+\!\rho)|m|-\rho m^2+(1\!+\!\rho)\Xi |m|(m^2\!-\!1)}\Big\},&&
\end{eqnarray}
where $m=\pm1,\pm 2,\pm 3, \dots$, and $\rho=\rho_{\rm in}/\rho_{\rm out}$.
For sufficiently small $\Xi<\Xi_{\rm c}(\rho)$, the radicand in Eq.(6) can be
negative. In this case, several modes with numbers 
$1<m_{\rm min}\leq |m|\leq m_{\rm max}$ are unstable. In particular,
$\Xi_{\rm c}(1)=1/15$ at equal densities, and the mode with $m=2$ remains stable.

Comparison of the conditions of both instabilities
demonstrates that there is a ``stability window'' $\Xi_{\rm c}(\rho)<\Xi<1$.

Note that the above dispersion law and the following 
condition for the stability of a conservative vortex
differ from those for a dissipative filled vortex considered 
in [34]. For stability, a dissipative vortex must be
near a strict minimum of the free energy and its
dynamics contains elements of ``gradient descent,''
whereas our vortex is described by Hamiltonian dynamics at the energy 
integral level that does not correspond to a static minimum.

The analogy between a binary condensate and a two-fluid 
classical system suggests that similar unstable 
and stable regimes should exist for a multiple vortex 
in the condensate, depending on the parameter 
$\sigma R/(\rho_{\rm out}Q^2)\sim \sqrt{g\mu} R/Q^2$. 
However, full agreement cannot be expected, at least because 
of the finite thickness of the transition layer.

\begin{figure}
\begin{center}
\epsfig{file=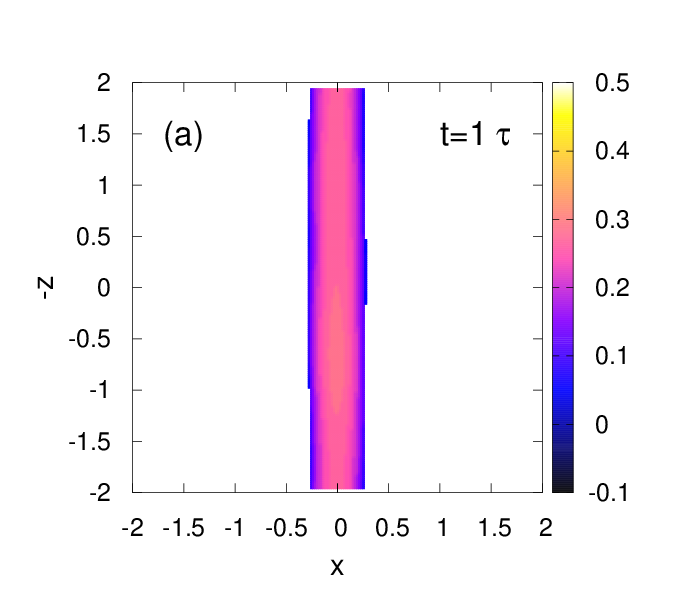, width=41mm}
\epsfig{file=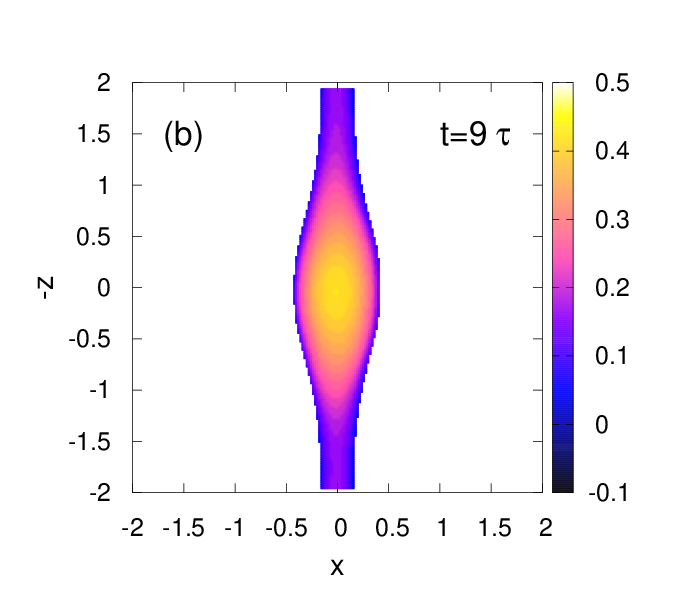, width=41mm}\\
\epsfig{file=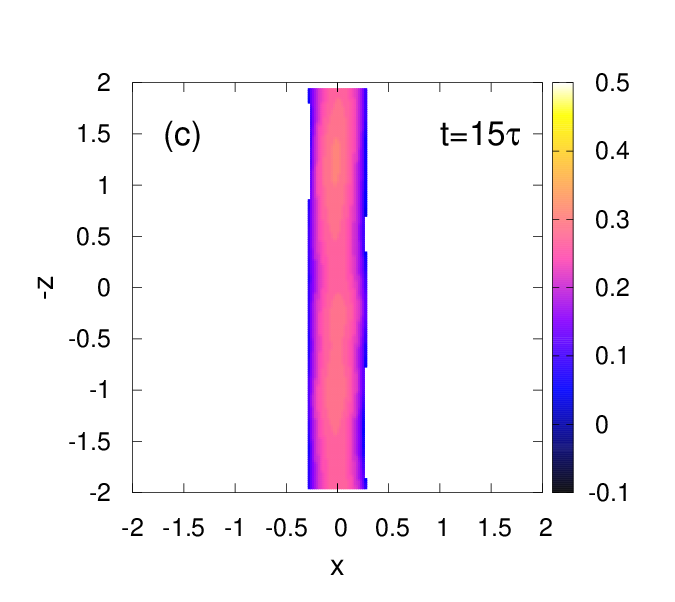, width=41mm}
\epsfig{file=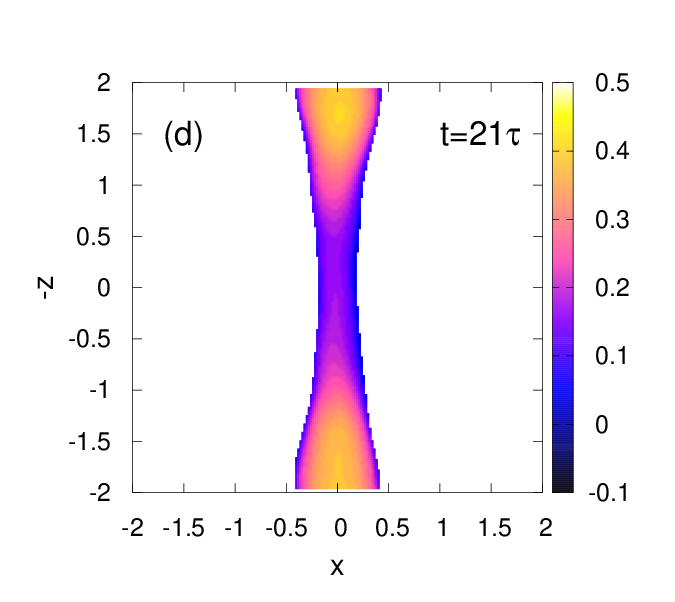, width=41mm}\\
\epsfig{file=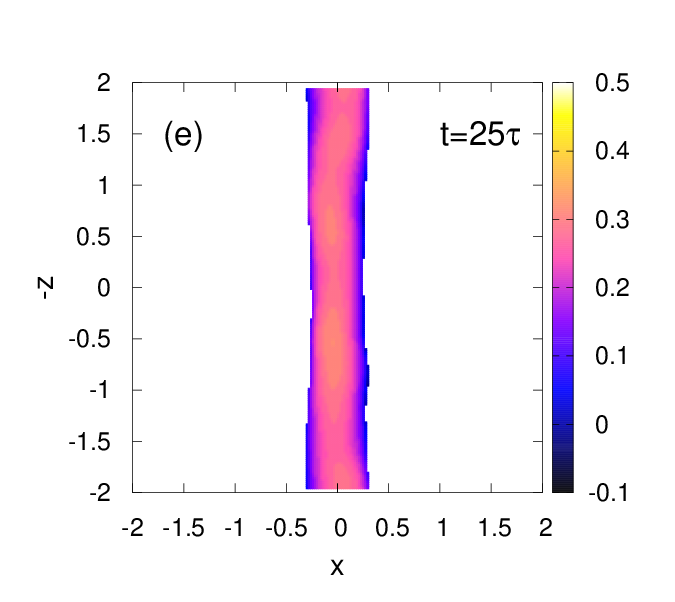, width=41mm}
\epsfig{file=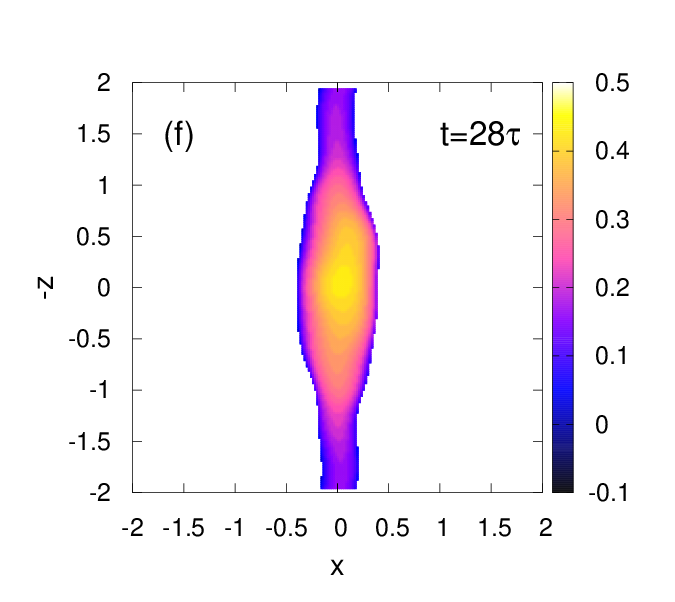, width=41mm}
\end{center}
\caption{ Example of the development of the sausage instability of a filled vortex 
and return to a weakly disturbed state. The color indicates the $y$ coordinate 
of the conditional interface. The parameters are $Q=1$, $g=0.05$, and $N_2/N_1=1.7/25.8$.
Distortion increases gradually, but at a lower rate.
}
\label{sausage_instability} 
\end{figure}
\begin{figure}
\begin{center}
\epsfig{file=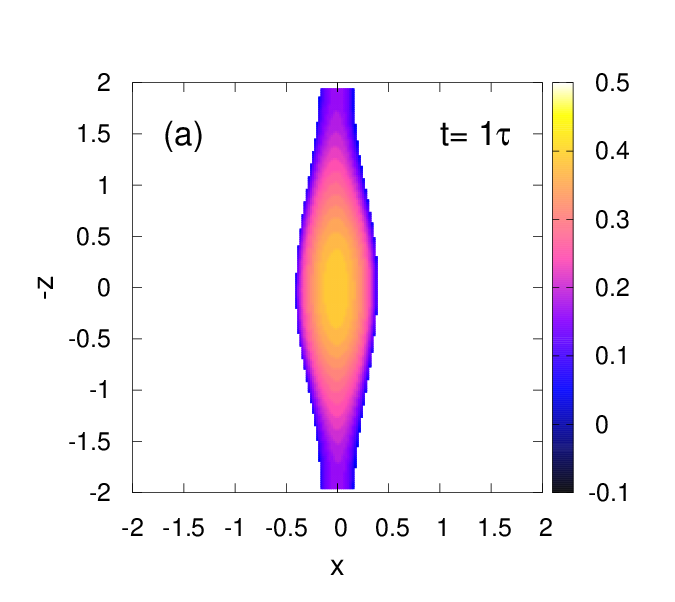, width=41mm}
\epsfig{file=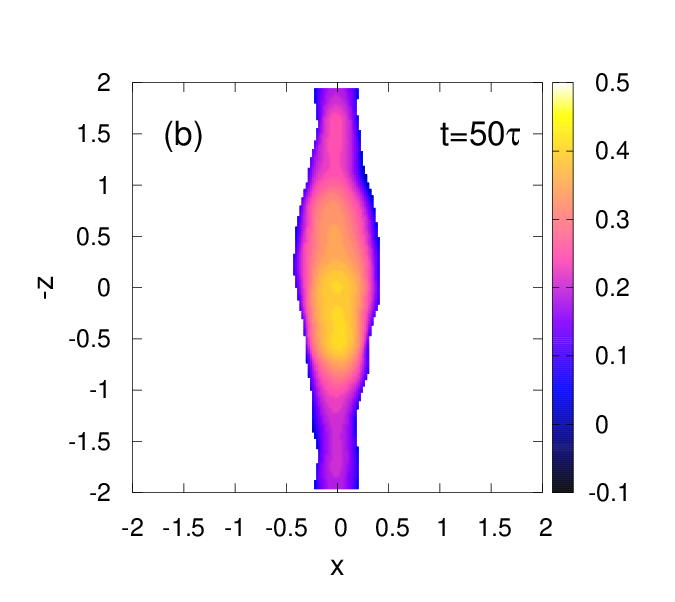, width=41mm}
\end{center}
\caption{ Example of an approximately stationary sausage structure. 
The parameters are $Q=1$, $g=0.05$, and $N_2/N_1=1.9/25.6$.}
\label{sausage_structure} 
\end{figure}
\begin{figure}
\begin{center}
\epsfig{file=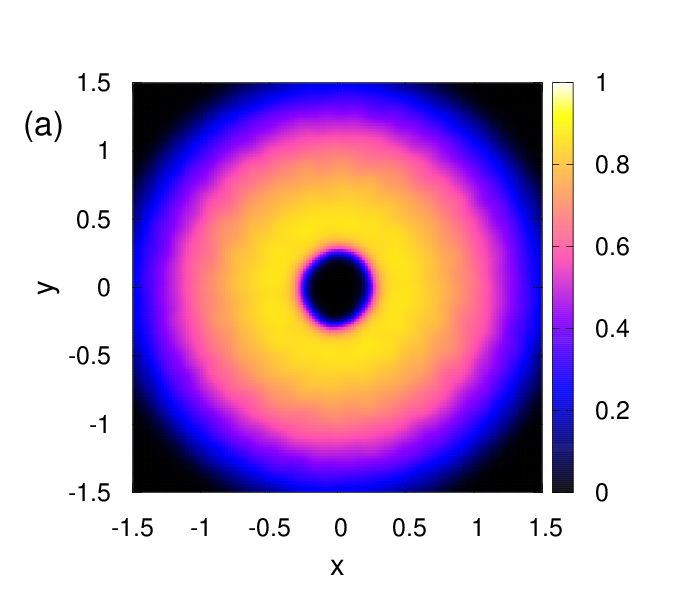, width=41mm}
\epsfig{file=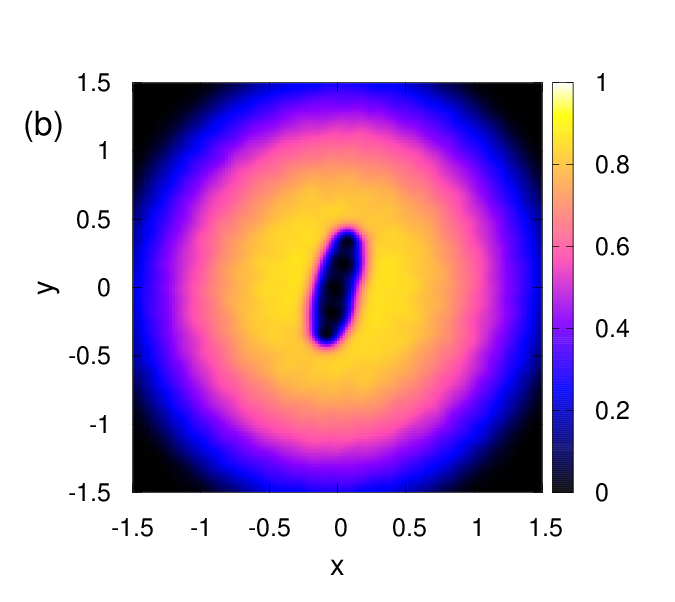, width=41mm}\\
\epsfig{file=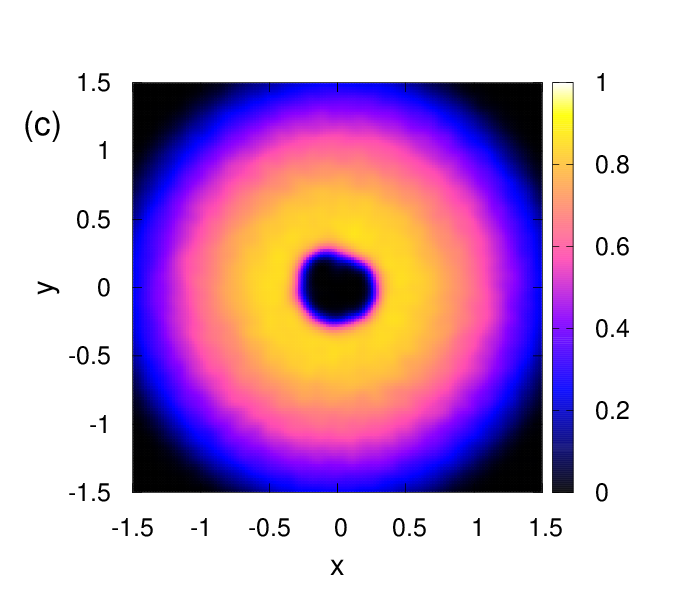, width=41mm}
\epsfig{file=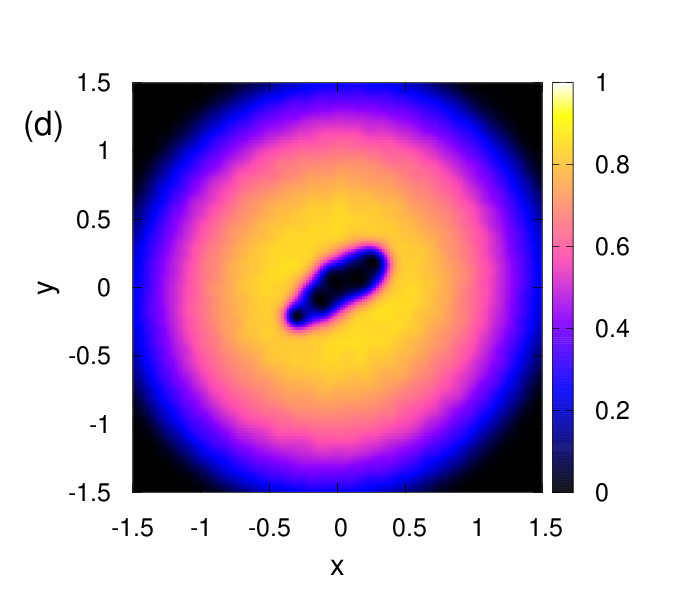, width=41mm}\\
\epsfig{file=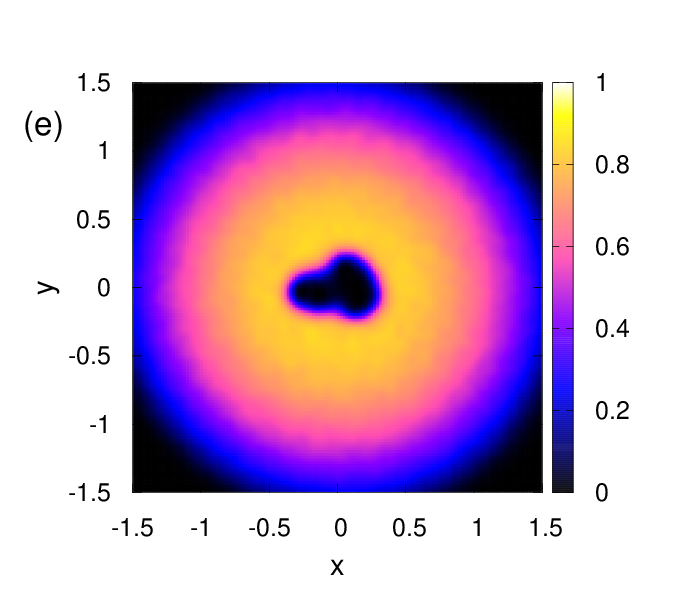, width=41mm}
\epsfig{file=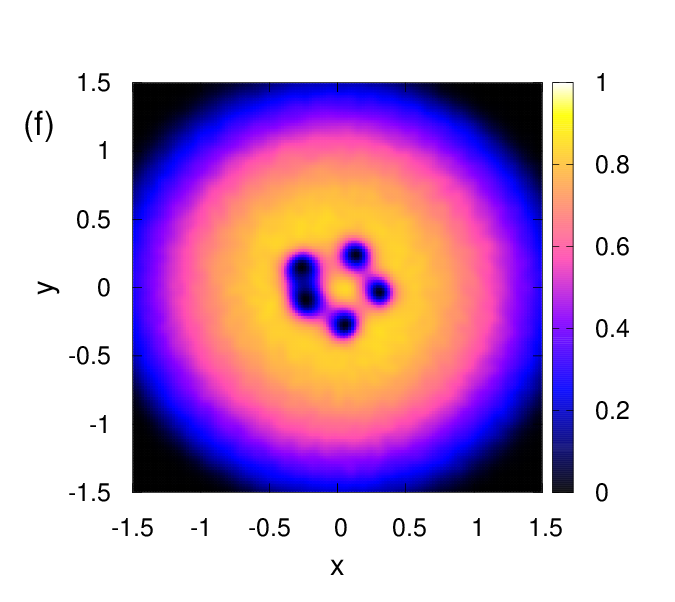, width=41mm}
\end{center}
\caption{Example of the development of the transverse instability of a filled vortex. 
The color indicates the rescaled density of the vortex component in the $z=0$ section 
at times $t$ = (a) $1\tau$, (b) $6\tau$, (c) $9\tau$, (d) $12\tau$, (e) $16\tau$, 
and (f) $28\tau$. The parameters are $Q=5$, $g=0.05$, and $N_2/N_1=1.0/24.8$.}
\label{2D_instability} 
\end{figure}

To confirm these considerations, I carried out a
numerical simulation of the system of evolutionary
equations (1) and (2) with the parameters $\nu=1$, $g_{11}=g_{22}=1$, 
and $g= 0.01, 0.02, 0.05, 0.10$. The split-step Fourier method 
and periodic boundary conditions in spatial coordinates were used. 
The calculation accuracy was controlled by obtaining the energy
integral and two integrals of the ``number of particles''
$N_1$ and $N_2$ up to the fifth decimal place.

Since it is rather difficult to study a homogeneous
condensate in a numerical experiment because of the
long-range nature of vortices, the external quadratic
potential $V_1=V_2=(x^2+y^2)/2$ was used. This led to
the practical transverse confinement of the condensate
in the size of $R_\perp=\sqrt{2\mu}$ and negated the interaction 
with the transverse boundary of the computational domain. 
A sufficiently large chemical potential $\mu=40$
provided the necessary conditions $w\ll R\ll R_\perp$. 
The equilibrium vortex radius $R$ was specified indirectly 
through the ratio of the number of particles $N_1$ and $N_2$
using a special numerical procedure which gives an 
approximately equilibrium initial vortex profile with small 
disturbances in the shape of the interface. Typical values were $R\sim 2$.

To exclude large values, the coordinates in Figs.1-5 
are rescaled to $R_\perp\rightarrow \sqrt{3}\approx 1.7$. 
The computational domain is a cube with a side of $2\pi/1.6\approx 4$. 
For the time scale, the number $\tau=2\mu/(3\cdot 2.56)\approx 10$ is used.
Wavefunctions are also rescaled: $(A,B)=\sqrt{\mu}(\psi,\psi_b)$ .
An equilibrium profile of the total density 
$(|\psi|^2+|\psi_b|^2)\approx [1-(x^2+y^2)/3]$ is obtained.

An example of the development of a moderate sausage 
instability is shown in Fig. 1. Owing to the inertia
of the process, the ``bubble'' on the vortex assembles
and decays several times. In other simulations, with
a larger value of $g$, the bubble was almost spherical and
then usually moved away from the axis of the system
and collapsed at the condensate boundary. This case is
not shown here.

If the initial configuration is set with a thickening
on the vortex (and with approximately zero poloidal
velocity), then such a bubble can retain its identity for
a long time, remaining in a relatively unchanged form
close to the extremal of functional (5) under the additional 
constraint $\int S dz = const$. However, three-dimensional 
distortions are gradually accumulating. A corresponding
example is shown in Fig. 2.

\begin{figure}
\begin{center}
\epsfig{file=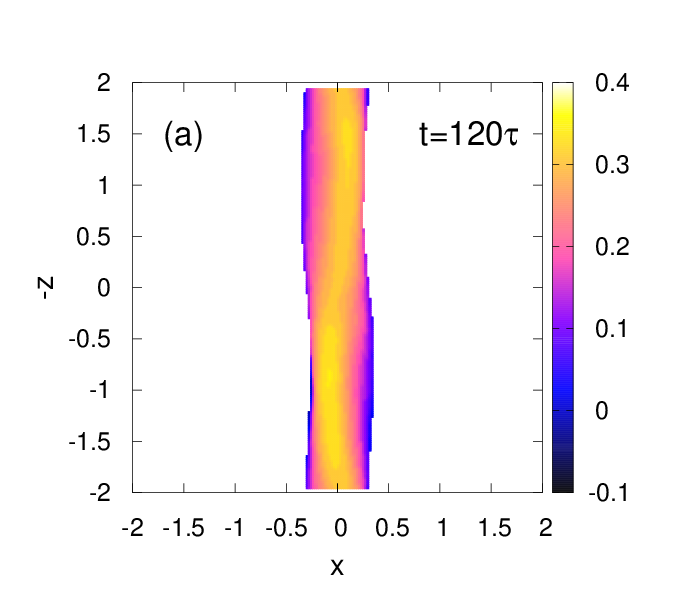, width=41mm}
\epsfig{file=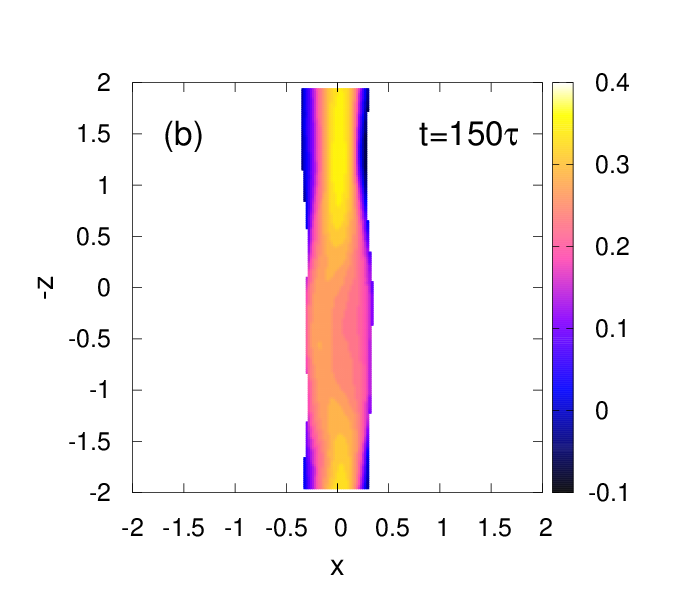, width=41mm}
\end{center}
\caption{Coherent structures at the nonlinear stage of the transverse instability. 
The parameters are $Q=5$, $g=0.05$, and $N_2/N_1=1.4/24.6$.}
\label{2D_structure} 
\end{figure}
\begin{figure}
\begin{center}
\epsfig{file=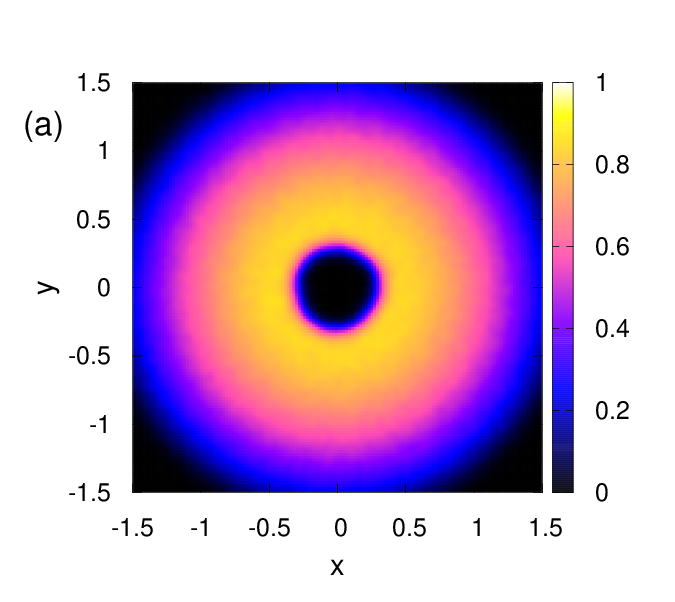, width=41mm}
\epsfig{file=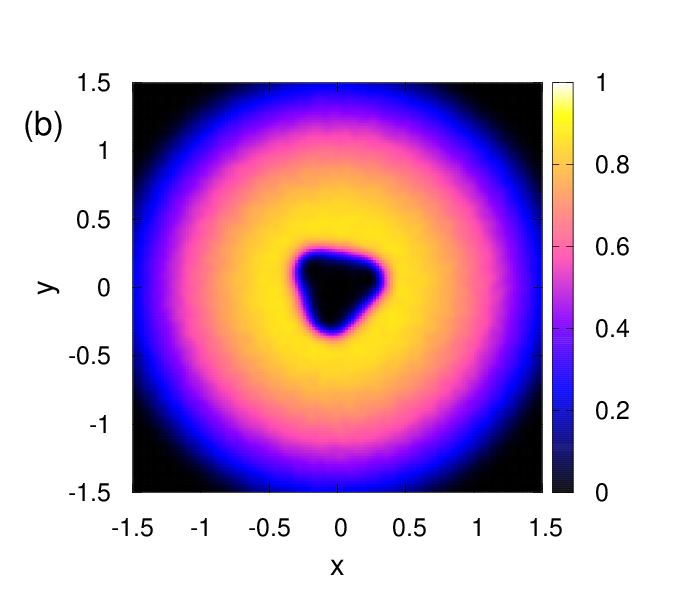, width=41mm}
\end{center}
\caption{Vortex cross sections in the $z=0$ plane corresponding 
to the times shown in Fig. 4: $t =$ (a) $120\tau$ and (b) $150\tau$.}
\label{2D_sections} 
\end{figure}

Figure 3 illustrates the development of the transverse
instability. In this example, the mode with $m=2$
is the most unstable, which contradicts the prediction
of the classical model. Apparently, the reason for this
is the moderate $R/w$ ratio (in this case, along with the
surface tension, the bending energy of the interface is
also included). The most significant difference from
the previous example is the use of a sufficiently large
vortex charge $Q=5$ in order to reduce the parameter
$\sqrt{g\mu} R/Q^2$. Initially, the cross section of the vortex
changes from round to elliptical; then, it becomes
approximately round again, further stretches, and
finally changes to an irregular shape. At the final stage,
the filled multiple vortex transforms into a cluster of
single filled vortices. For comparison, in the case with
$Q=4$ (not presented here) and approximately the
same radius $R$, the dynamics remained stable. However, 
with a decrease in the segregation parameter to
$g=0.01$ and a lower filling of the core $N_2/N_1\approx 0.7/25.7$, 
the instability developed according to a qualitatively 
similar scenario at $Q=4$.

Note that only multiple filled vortices enter the stability region, 
while multiple vortices in a one-component condensate are unstable.

At a lower level of supercriticality, the transverse
instability can lead to the spontaneous formation of
long-lived three-dimensional coherent structures.
Figures 4 and 5 show the results of a numerical experiment 
where the filling of the vortex with the bright
component was increased in comparison with Fig. 3.
In particular, the critical parameter increased and
approached the edge of the unstable region, but
remained in it. The mode with $m=3$ was the most
unstable, and as a result, most of the vortex section
became a rounded triangle. The parameters of this
section (angle of rotation minus homogeneous uniform 
rotation and deviation from the circular shape)
depended on the time and longitudinal coordinate. A
distant analogy of such vortices with nonaxisymmetric
vortices in superfluid $^3$He-$B$ can be noted.

To summarize, a critical parameter that qualitatively 
determines stable and unstable regimes in the
dynamics of a highly filled quantum vortex in a binary
segregated Bose-Einstein condensate has been proposed 
in this work. Numerical examples are given for
unstable regimes, including those with long-lived
three-dimensional coherent structures.

\end{document}